# Efficient high harmonic generation in nonlinear photonic moiré superlattice

Tingyin Ning,[1,2*] Yingying Ren, [1,2] Yanyan Huo,[1,2] and Yangjian Cai[1,2]

[1]Shandong Provincial Engineering and Technical Center of Light Manipulations & Shandong Provincial Key Laboratory of Optics and Photonic Device, School of Physics and Electronics, Shandong Normal University, Jinan 250358, China

[2]Collaborative Innovation Center of Light Manipulation and Applications, Shandong Normal University, Jinan 250358, China

*Corresponding author: ningtingyin@sdnu.edu.cn

**Abstract:** Photonic moiré superlattice as an emerging platform of flatbands can tightly confine the light inside the cavity and has important applications not only in linear optics but also in nonlinear optics. In this paper, we numerically investigate the third- and fifth-order harmonic generation (THG and FHG) in photonic moiré superlattices fabricated by the nonlinear material silicon. The high conversion efficiency of THG and FHG is obtained at a relatively low intensity of fundamental light, e.g., the maximum conversion efficiency of THG and FHG arrives even up to be $10^{-2}$ and $10^{-9}$ at the fundamental intensity of 30 kW/m$^2$, respectively, in the moiré superlattice of near flat band formed by the twist angle 6.01°. The results indicate the photonic moiré superlattice of a high-quality factor and flatbands is a promising platform for efficient nonlinear processes and advanced photonic devices.

## 1. Introduction

Harmonic generation is one of the fundamental nonlinear optical processes, and has attracted endless interests since the birth of nonlinear optics from the second harmonic generation (SHG) [1,2]. Frequency conversion, as one of the most important applications of harmonic generation, is practically used to produce a short-wavelength lasing, and thus extends laser sources from near-infrared to visible and violate regions [1]. Traditionally, the efficient harmonic generation in bulk materials, such as nonlinear crystals or waveguides, requires the phase matching between the fundamental and harmonic generation beams [1]. However, the integrated nanophotonic devices require the reduced size of the structures down to the subwavelength domain. The phase matching is no longer considered as the way to improve the efficiency of nonlinear responses. On the other hand, enhancing the local field in the subwavelength resonant structures becomes an alternative and effective method for the efficient harmonic generation [3-5].

Different types of subwavelength structures of enhanced local fields for efficient nonlinear processes have been widely considered. Metal nanostructures have received increasing attention to enhance the nonlinear response since 1980's [6]. The local field around the metal surface is greatly enhanced due to the excitation of surface plasmon resonance. With the development of the technique of nanofabrication and chemical synthesis, various shapes and arrangements of metallic nanostructures were investigated for efficient nonlinear responses and phase controls [3,7-10]. However, the efficiency of nonlinear responses in such plasmonic systems remains relatively low due to the surface nonlinearity, large losses and the low damage threshold of metallic nanostructures [3]. Nanostructures made of dielectric materials of a high-refractive-index as emerging platforms for nonlinear optics have drawn more attentions due to the large bulk nonlinear susceptibility, negligible loss, and high damage threshold [4,5]. Large nonlinear optics responses in various all-dielectric resonators have been widely reported in recent years, including photonic crystals [11-13], resonant waveguide gratings [14-16], microrings [17-19], metasurfaces [20-22]. Harmonic generation in all dielectric resonators was significantly enhanced. The typical SHG conversion efficiency in dielectric resonators reached up to $8\times10^{-6}$ when the intensity of fundamental light, $I_{FF}$, is at around 1 GW/cm$^2$ level [23], and even $10^{-3}$ at the $I_{FF}$ of 1.6 GW/cm$^2$ [24]. THG conversion efficiency in silicon resonators arrived to be $10^{-6}$ at the $I_{FF}$ of 3.2 GW/cm$^2$ [25], and even $10^{-4}$ at the $I_{FF}$ of 3 GW/cm$^2$ [26]. Particularly, photonic bound states in the continuum (BICs), a concept of the existence of confined states embedded in the continuum spectrum, can perfectly confine the light in a nanostructure without any radiation of an infinite quality($Q$)-factor. In recent years, BICs or quasi-BICs (QBICs) were intensively studied to enhance the harmonic generation [27]. SHG conversion efficiency in all-dielectric nanostructures boosted by (Q)BICs arrived up to $10^{-3}$ at the $I_{FF}$ of kW/cm$^2$ level [28] and even $10^{-1}$ at the $I_{FF}$ of MW/cm$^2$ level [29], and the typical THG conversion efficiency observed experimentally was around $10^{-6}$ at the $I_{FF}$ of 0.1 GW/cm$^2$ [30]. Especially, the efficient high-harmonic generation (HHG) process, e.g., fifth-order harmonic generation, has attract interesting and be obtained in such high $Q$-factor resonators of BICs [31-33], and even up to 11$_{th}$-order was experimentally observed due to the BICs in silicon metasurfaces [34].

Moiré superlattices have given birth to many novel physical phenomena in electronics [35-37]. Photonic moiré superlattice as an emerging platform have also drawn much attention, since they can extremely confine the light field to enhance the light-matter interaction [38,39]. Especially, the flatband modes in moiré superlattices can break the limitation of dispersive effects in traditional resonators, such as quasi-BICs in photonic crystals, resonant waveguide gratings, which are highly dependent on the wavevector *k*. Such the *k*-independent dispersionless flatband in the moiré superlattice enables it to work under wide-angle illuminations. Many novel photonic applications, including low-threshold lasing [40], all-optical switching empowered by ultra-low intensity [41], wide-angle absorbers and reflectors [42], enhanced SHG efficiency in $WS_2$ monolayer placed on the moiré superlattice [43], and tunable harmonic generations in twisted 2D materials [44-46], were investigated recently, indicating the photonic moiré superlattice can be used as a promising platform of photonic devices .

In this article, we report the efficient HHG, including THG and FHG, in the silicon photonic moiré superlattice. The large conversion efficiency is obtained at relative low intensities in moiré superlattices of different commensurate twist angles. The results open new possibilities for efficient harmonic generations based on photonic moiré superlattices.

**2. Moiré superlattice and numerical modeling**

Moiré superlattices are traditionally formed by twisting two periodical patterns with a particular angle or shifting two periodical patterns of a mismatched lattice constant. In this article, we consider the moiré superlattice formed by twisting two layers of triangular lattices in the same silicon membrane, which has been proved to be fabricated using state-of-the-art nanofabrication techniques [40]. A relatively large twist angle $\theta$ of 13.17º is first chosen to optimize the geometry. Comparing with the smaller commensurate magic angles, such as 9.43º, 2.65º, the twist angle 13.17º produces the relative smaller unit cell of the superlattice to save the computation time. The schematic structure of the moiré superlattice is shown in Fig. 1 (a). The lattice constant *a* is determined by the lattice constant of the basic triangular lattice $a_0$ and the twist angle $\theta$ as $a=a_0/(2\sin(\theta/2))$. The optimized parameters $a_0 = 600$ nm and $r_0 = 100$ nm are used, where $r_0$ is the radius of holes in the silicon slab of the thickness 200 nm.

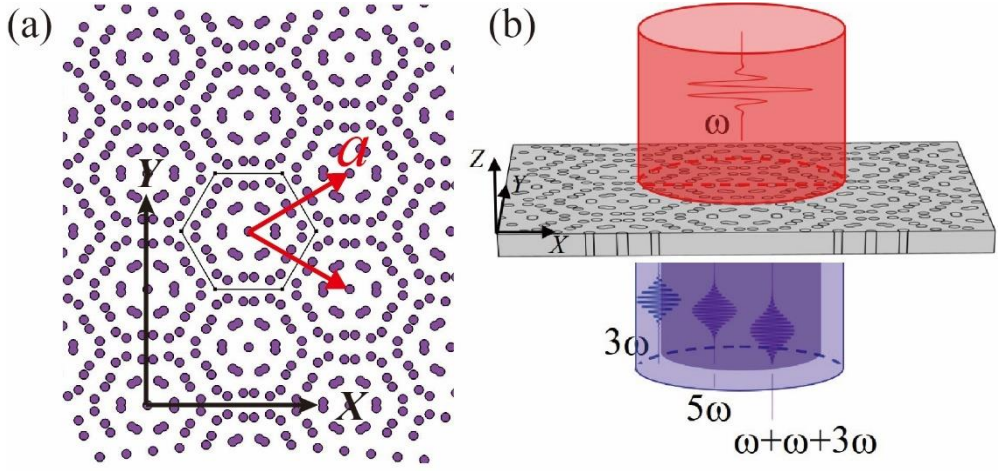

Fig. 1. (a) Schematic structure of a moiré superlattice of lattice constant $a$. (b) The process of THG and FHG at the fundamental frequency ω.

The fundamental light of angular frequency $\omega$ illuminates on the nanostructure at the normal of incidence, as shown in Fig. 1(b). Due to the symmetry of silicon structure, the even-order harmonic generation is forbidden. The THG and FHG in the moiré superlattice are considered. Especially, the direct process of FHG, which is a process of five photons of fundamental frequency $\omega$ converted into one photon of frequency $5\omega$, and the cascaded process of FHG, which is another process via the degenerate four-wave mixing between two fundamental photons and one THG photon, are both taken into account. The governing wave equations are similar as we reported in Ref. [47]. The nonzero tensor components of the third-order nonlinear susceptibility are $\chi^{(3)}_{iiii}$, $\chi^{(3)}_{iijj}$, $\chi^{(3)}_{ijij}$, and $\chi^{(3)}_{ijji}$, where $i, j$ represents $x$, $y$ and $z$. The $\chi^{(3)} = \chi^{(3)}_{iijj} = \chi^{(3)}_{ijij} = \chi^{(3)}_{ijji} = 1/3\,\chi^{(3)}_{iiii}$ is assumed, and $\chi^{(3)}_{iiii} = 2.45\times10^{-19}$ m$^2$/V$^2$ [48]. Similarly, the dominant nonzero tensor components of the fifth-order nonlinear susceptibility is $\chi^{(5)}_{iiiii}$, and the other components with subscripts of both *i* and *j*, such as $\chi^{(5)}_{iiijji}$, $\chi^{(5)}_{iijiji}$, $\chi^{(5)}_{ijjiii}$, $\chi^{(5)}_{ijijii}$, $\chi^{(5)}_{ijjiij}$, $\chi^{(5)}_{iijijj}$, and $\chi^{(5)}_{ijijij}$, are assumed to be equal to each other of the value one-third of $\chi^{(5)}_{iiiii}$ and denoted as $\chi^{(5)}$. The $\chi^{(5)}_{iiiii}$ is estimated to be $5.6\times10^{-19}$ m$^4$/V$^4$ based on the phenomenological relation, $\chi^{(n)} \approx \chi^{(1)}/E_{at}^{(n-1)}$, where $\chi^{(1)}$ is the linear susceptibility and $E_{at}$ is the atomic field strength [1,31]. Thus, the THG polarization **P**(3ω) is expressed as $\mathbf{P}(3\omega) = 3\chi^{(3)}(\mathbf{E}(\omega)\cdot\mathbf{E}(\omega))^2\mathbf{E}(\omega)$, and the FHG of the direct process and cascaded process is written as $\mathbf{P}_d(5\omega) = 3\chi^{(5)}(\mathbf{E}(\omega)\cdot\mathbf{E}(\omega))^2\mathbf{E}(\omega)$, and $\mathbf{P}_c(5\omega) = 3\chi^{(3)}[2(\mathbf{E}(\omega)\cdot\mathbf{E}(3\omega))\mathbf{E}(\omega) + (\mathbf{E}(\omega)\cdot\mathbf{E}(\omega))\mathbf{E}(3\omega)]$, respectively. In addition, since the conversion efficiency of THG is some large at certain fundamental intensities, the mixing of fundamental and THG waves is not negligible, and thus such the nonlinear polarization at ω, **P**$^{NL}$(ω), should be $\mathbf{P}^{NL}(\omega) = 3\chi^{(3)}(\mathbf{E}(3\omega)\cdot\mathbf{E}^*(\omega))\mathbf{E}^*(\omega)$, which was usually neglected in the previous investigations [31]. The coupled wave equations can be numerical solved using finite elements methods (FEM) by the commercial software Comsol Multiphysics in the frequency domain. Four steps are needed for linear and nonlinear responses here. The fundamental field of frequency ω is calculated in the step one. A plane wave of a polarization along the *x*-axis is introduced by one port, and propagates through the moiré superlattices to the other port. The periodic conditions are employed on the surrounding boundaries, and the vector for Floquet periodicity $k_F$ is set as 0 due to the normal incidence. The fundamental field E(ω) and linear transmittance can be obtained in this step if pure linear optical response is considered. The THG response in the moiré superlattices is calculated in the step two. In this step, the fundamental field E(ω) in the step one is used to express the components of THG polarization **P**(3ω) of silicon. It is noting that the step one and step two are not independent but coupled, since the nonlinear polarization $\mathbf{P}^{NL}(\omega)$ of silicon, which involves E(3ω), should be included in the step one. So the step one and step two should be solved together to obtain the linear and nonlinear response. The power of generated THG signal can be calculated by the integration of Poynting flow at 3ω on the port one and the port two. The FHG via the direct and cascaded processes can be calculated separately in the step three and the step four. The polarizations $P_d$ (5ω) and $P_c$ (5ω) are directly written in the corresponding step based on the E(ω) and E(3ω) obtained in the step one and the step two. The power of FHG signals is also obtained by integrating of Poynting flow at 5ω on the port one and the port two. The refractive index of silicon at the concerned fundamental, THG and FHG wavelength region is set as $n(\omega)=3.48$, $n(3\omega)=3.9+0.022i$, and $n(5\omega)=6.66+1.66i$, respectively [49,50]. The conversion efficiency of THG and FHG is defined as $\eta_{THG}=P_{THG}/P_{FF}$, and $\eta_{FHG}=P_{FHG}/P_{FF}$, respectively, where $P_{FF}$ is the power at the fundamental frequency, and $P_{THG}$ and $P_{FHG}$ are the power of THG and FHG, respectively.

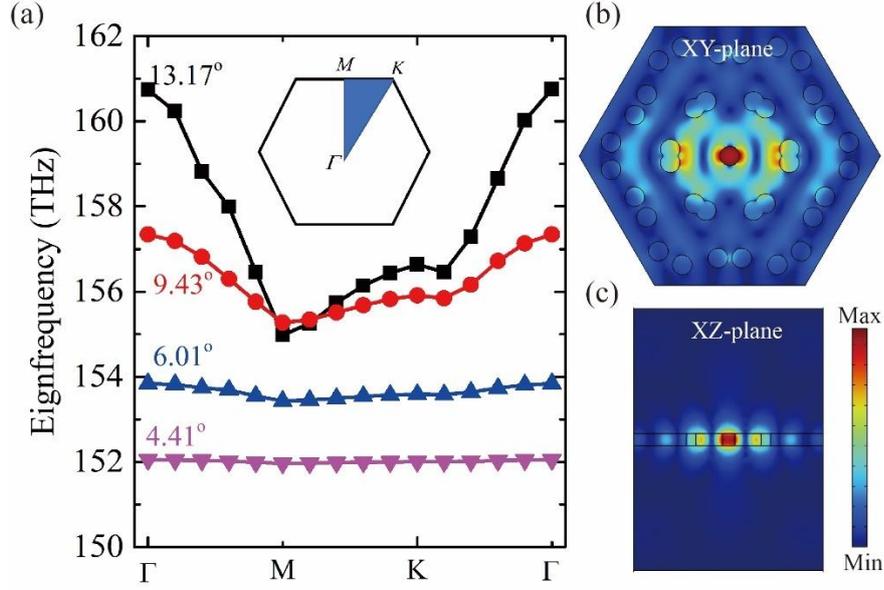

Fig. 2. (a) The band diagrams of the $p_x$ mode under different commensurate twist angles, (b) and (c) show the electric field distribution at the $p_x$ mode in the moiré superlattice of the twist angle 13.17°.

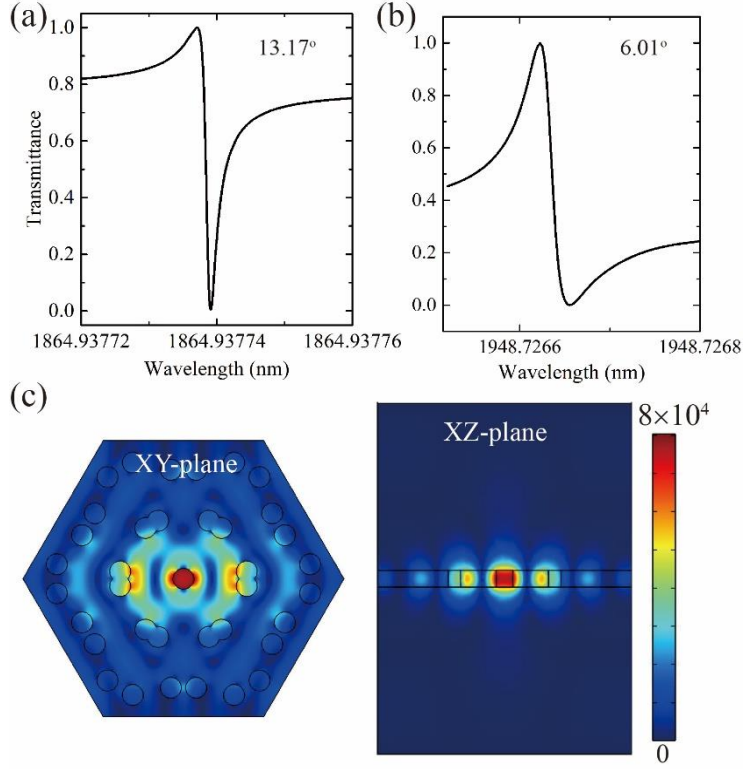

Fig. 3. The spectrum of linear transmittance from the moiré superlattices formed by the twist angle (a) 13.17° and (b) 6.01°. (c) The distribution of electric field at the resonance wavelength, i.e. $p_x$ mode, at the XY- and XZ-plane, respectively. The twist angle is 13.17°.

The eigenfrequency analysis was first conducted by Comsol Multiphysics to determine the resonance modes. Conventionally, there are two near-degenerate modes $p_x$ and $p_y$ with the electric field confined at the center of the moiré superlattice which are mostly concerned, since they can form the flatband in the moiré superlattices of small twisted angles [40]. In this article, we only concern the $p_x$ mode, and the optical behaviors at the $p_y$ mode are similar as those at the $p_x$ mode just of different values. The band diagrams of the $p_x$ mode in moiré superlattices of different commensurate twist angles are shown in Fig. 2(a). The eigenfrequency become smaller with the decrease of twist angles, which indicates that the resonance modes can be dramatically tuned by the twist angle. The near flat band is formed when the twist angle reduces to be smaller than 6.01°. The eigenfrequency of $p_x$ mode at the Γ point in the moiré superlattice of the twist angle 13.17° is obtained about 160.752 THz, and the distribution of electric field at the XY-plane and XZ-plane is shown in Fig. 2(b) and (c), respectively.

## 3. Results and discussion

The linear transmittance of nanostructured moiré superlattices of different twist angles is calculated. Fig. 2(a) and (b) present the transmittance from the moiré superlattices of the twist angle 13.17° and 6.01°, respectively. The resonance wavelength is around 1864.93774 nm and 1948.72665 nm when the twist angle is 13.17° and 6.01°, respectively, which correspond to the $p_x$ mode obtained by the eigenfrequency analysis. The Q-factor of such the Fano lineshape is calculated by $Q=\lambda_r/\Delta\lambda$, where $\lambda_r$ and $\Delta\lambda$ are the resonance wavelength and the difference between the peak and dip, respectively. The value of Q-factor is around $8.5\times10^8$ and $6\times10^7$ in the moiré superlattices of the twist angle 13.17° and 6.01°, respectively. The reduced Q-factor in the nanostructure of the twist angle 6.01° is ascribed to the unoptimized geometry. The high Q-factor is responsible for the efficient HHG in the moiré superlattices. The local electric field distributions in the nanostructure of the twist angle 13.17° are shown in Fig. 3(c). The enhancement factor of local field, $|\mathbf{E}(\omega)/\mathbf{E}_{in}(\omega)|$ with $\mathbf{E}_{in}(\omega)$ the incident electric field of 1 V/m along the x-axis, arrives up to $8\times10^4$.

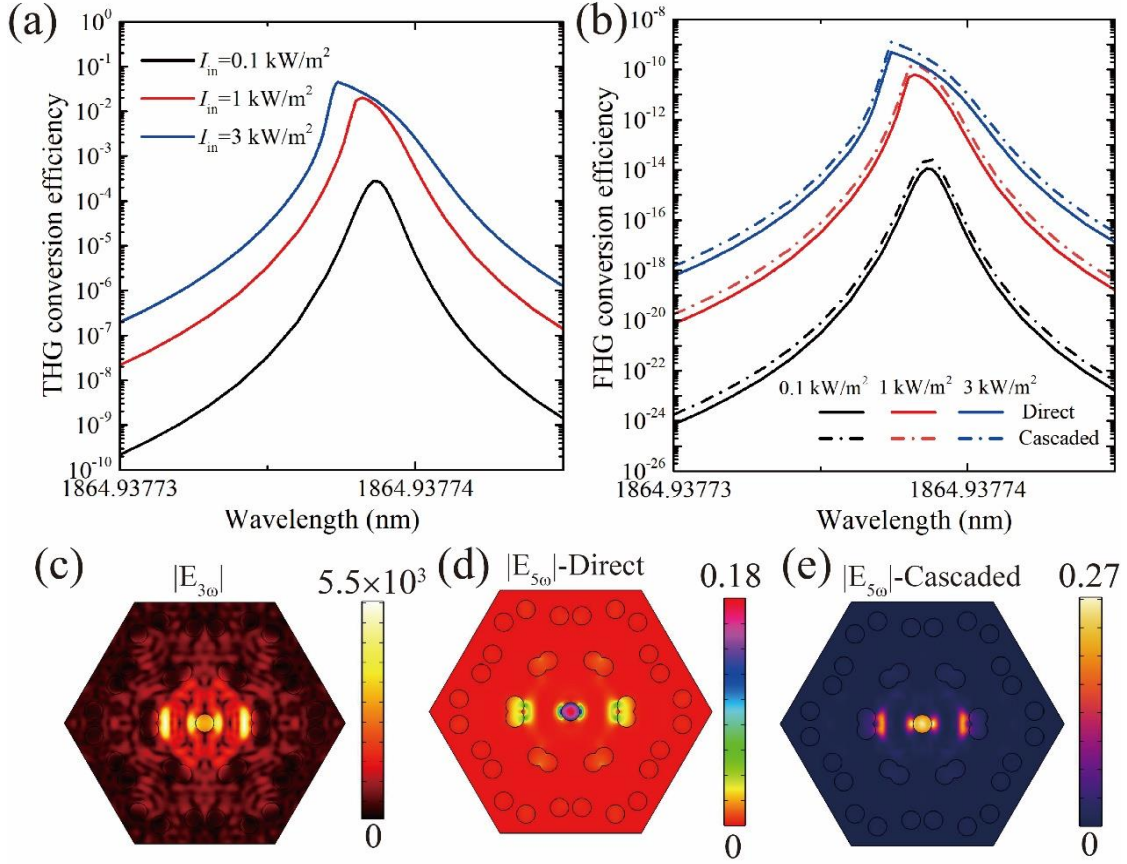

Fig. 4. (a) and (b) Dependence of conversion efficiency of THG and FHG in the moiré superlattice of the twist angle 13.17° on the fundamental wavelength at different intensities of fundamental light. (c)-(e) The electric field distribution of THG, direct FHG and cascaded FHG at the $p_x$ mode under the fundamental intensity 1kW/m$^2$.

The dependence of conversion efficiency of THG and FHG in the moiré superlattice of the twist angle 13.17° on the fundamental wavelength at different pump intensities is shown in Fig. 4(a) and (b), respectively. It is clear that the maximum $\eta_{THG}$ and $\eta_{FHG}$ occurs at the resonance wavelength of $p_x$ mode at a relatively low fundamental intensity, e.g. 0.1 kW/m$^2$. At such intensity, the maximum $\eta_{THG}$ is around 10$^{-4}$, which indicates that the THG signal does not affect the polarization of $\mathbf{P}^{NL}(\omega)$, and can be neglected safely. However, with the increase of fundamental intensity to 1 kW/m$^2$, the $\eta_{THG}$ becomes larger up to 10$^{-2}$, and a clear shift of the maximum value from the original resonance fundamental wavelength at the $p_x$ mode is observed. Such shift of the resonance mode is ascribed to the nonlinear polarization $\mathbf{P}^{NL}(\omega)$ due to the wave mixing of fundamental and THG waves. Further increasing the fundamental intensity, the shift of the maximum $\eta_{THG}$ is clearer. The maximum $\eta_{THG}$ reaches up to $10^{-1}$ at the fundamental intensity of 3 kW/cm$^2$. The $\eta_{FHG}$ has the same trend as the $\eta_{THG}$. The $\eta_{FHG}$ is too low to affect the polarization of $\mathbf{P}^{NL}(\omega)$, and thus it is safely neglected and only the wave mixing between fundamental and THG waves was considered. It is also noted that the $\eta_{FHG}$ of the cascaded FHG is higher than that of the direct FHG, due to the relation between the $\chi^{(5)}$ and $\chi^{(3)}$, as interpreted in Ref. [30]. The maximum cascaded $\eta_{FHG}$ arrives up to $10^{-10}$, and even $10^{-9}$ at the fundamental intensity 1 kW/m$^2$ and 3 kW/m$^2$, respectively. The field distributions at THG and FHG at the fundamental intensity 1 kW/m$^2$ are shown in Fig. 4 (c)-(e).

When the fundamental wavelength is fixed at the $p_x$ mode, the dependence of $\eta_{THG}$ and $\eta_{FHG}$ on the intensity of fundamental light in the moiré superlattice of the twist angle 13.17° and 6.01° is given in Fig. 5 (a) and (b), respectively. The slopes of $\log_{10}(\eta_{THG})$ *vs* $\log_{10}(I_{in})$ and $\log_{10}(\eta_{FHG})$ *vs* $\log_{10}(I_{in})$ should be 2 and 4, since the $\eta_{THG}=P_{THG}/P_{FF}\sim (P_{FF})^2$ and $\eta_{FHG}=P_{FHG}/P_{FF}\sim (P_{FF})^4$. The slopes follow the predicated values well when the pump intensity is relatively low. The values of $\eta_{THG}$ and $\eta_{FHG}$ in the moiré superlattice of the twist angle 13.17° start to deviate from the predicted slopes when the intensity increases to be 1 kW/m$^2$. And such the threshold of pump intensity is around 10 kW/m$^2$ in the moiré superlattice of the twist angle 6.01°. The deviations are ascribed to the reduced fundamental field in the nanostructure as the shift of resonance mode due to the nonlinear polarization $\mathbf{P}^{NL}(\omega)$ (as shown in Fig. 4(a) and (b)). The different thresholds of the pump intensity for the slope deviation in moiré superlattices of different twist angles are due to the difference local field enhancement and the related THG efficiency. The larger local field enhancement and THG conversion efficiency will produce the larger $\mathbf{P}^{NL}(\omega)$ at a relatively lower pump intensity to shift the resonance mode. The conversion efficiency of THG and FHG in moiré superlattices of different twist angles when the fundamental wavelengths are set at their corresponding $p_x$ modes is shown in Fig. 6. When the pump intensity is lower than 1 kW/m², the conversion efficiency of THG and FHG has not deviated from the predicted slopes, and the efficiency of nonlinear response is determined by the local field enhancement. While when the pump intensity arrives or is larger than the threshold of pump intensity for the deviation, the efficiency of nonlinear response becomes complex and can not be predicted from the predicted slopes. When the pump intensity is 10 kW/m², the $\eta_{THG}$ in the moiré superlattice of the twist angle 13.17° is almost unchanged compared with that at the $I_{in}$=1 kW/m$^2$, while $\eta_{THG}$ in the moiré superlattice of the twist angle 6.01° is enhanced around two orders of magnitude compared with that at the $I_{in}$=1 kW/m$^2$. It is noting that although the geometry is not optimized for the moiré superlattice of the twist angle 6.01°, the conversion efficiency of THG and FHG in such superlattice of near flatband is still very high, especially considering the low input intensity. The further enhanced nonlinear responses should be arrived in the optimized moiré superlattice of the flatbands formed by small twist angles.

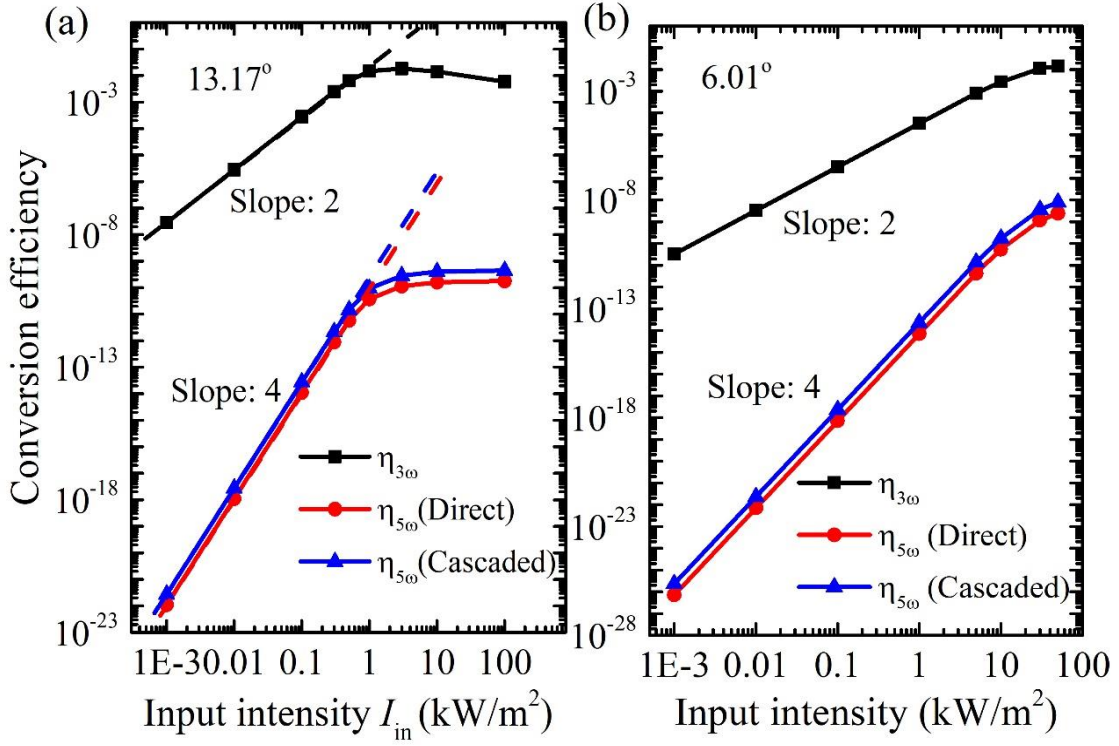

Fig. 5. Dependence of conversion efficiency of THG and FHG in the nanostructures of the twist angle (a) 13.17° and (b) 6.01° on the input intensity when the fundamental wavelength is fixed at the $p_x$ mode, respectively.

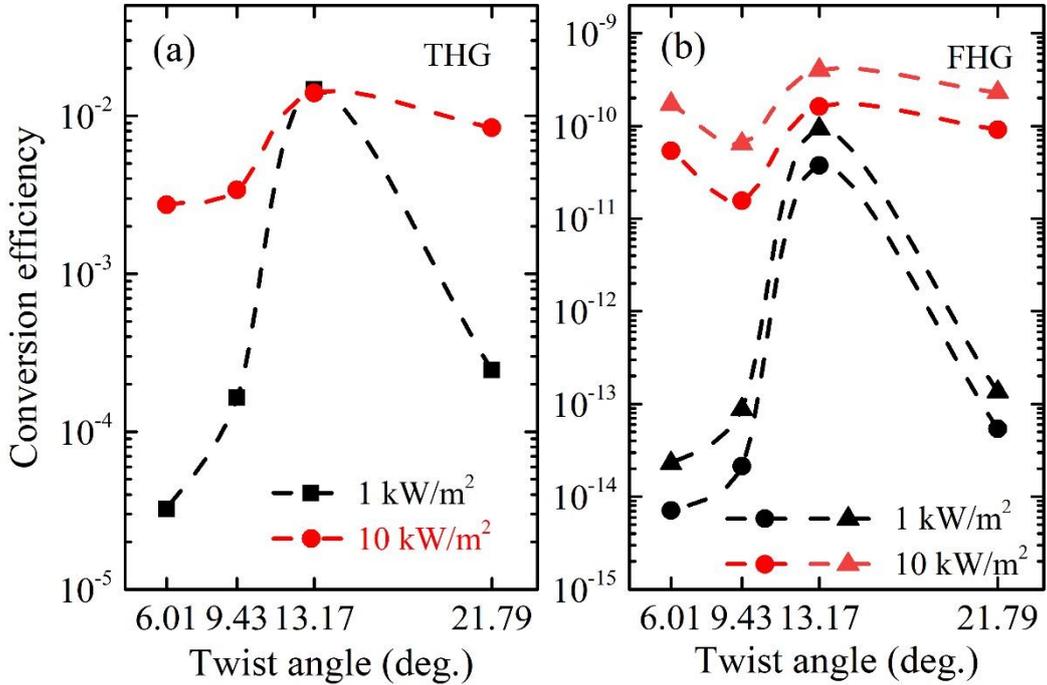

Fig. 6. Dependence of conversion efficiency of (a) THG and (b) FHG on the commensurate twist angles at the input intensity 1 kW/m$^2$ and 10 kW/cm$^2$, respectively.

Here, we only focus on the THG and FHG, but the higher order harmonic generations would be expected to have large conversion efficiencies in the nanostructured moiré superlattices. Not silicon but when the other nonlinear materials without the symmetric structure possessing both even- and odd-order nonlinear susceptibilities, such as AlGaAs, LiNbO$_3$, are employed for moiré superlattices, the supercontinuum generation is expected at the reasonable fundamental intensity.

Further, the moiré superlattice of a perfect flatband will be immune to the angle of incidence to produce efficient HHG or supercontinuum generation with a tightly focused beam or wide-angle sources [51]. In addition, the Kerr effect, i.e., the intensity-dependent refractive index, of silicon is not considered. Though the value of nonlinear refractive index of silicon at the fundamental wavelength is unknown, the Kerr effect of silicon would probably affect the resonance mode of nanostructure, especially when the intensity of fundamental light is some large. This is harmful to the efficiency of HHG. Finally, we would like to point out that the line shape of transmission spectrum will be broaden in the practical experiments due to the defects introduced by nanofabrication and the defects in the nonlinear material itself. The $Q$-factor and the enhancement factor of the local electric field, along with the efficiency of nonlinear response, will thus be decreased. Though the low threshold of nano-lasing in moiré superlattices has been experimentally realized, the fabricated moiré superlattice of an ultra-high $Q$ factor is still a challenge.

## 4. Conclusion

We numerically investigate the THG and FHG in photonic moiré superlattices fabricated by silicon. The high conversion efficiencies of THG and FHG are obtained at relatively low intensities of fundamental light. The efficient even- and odd-harmonic generation, and the higher harmonic generations other than FHG, are expected in nanostructured moiré superlattices of flatbands fabricated by nonlinear materials of broken symmetry structures. The mixing of efficient HHGs would finally give a supercontinuum generation. The results open a new possibility for efficient nonlinear process based on photonic moiré superlattices working in a broadband, at an ultra-low intensity and under wide-angle illuminations.

**Acknowledgements**

We are grateful for the financial support from National Natural Science Foundation of China (12174228, 12274271, 11874243).